\documentclass[epsfig,indentfirst,bm,12pt]{article}
\usepackage{CJK}
\usepackage{graphicx,amsmath,dcolumn,bm}
\usepackage{color}
\begin{document}
\title{\Large{\bf{The Drell-Yan nuclear modification due to the nuclear effects of nPDFs and initial-state parton energy loss}}
\thanks {Supported partially by National Natural Science
Foundation of China (11405043), Natural Science Foundation of Hebei Province (A2018209269) and Science and Technology Foundation of Hebei Education Department(ZD2020104).}}

\begin{CJK*}{GBK}{song}
\author{Li-Hua Song $^{1}$
\footnote{\tt{ E-mail:songlh@ncst.edu.cn}}
Peng-Qi Wang $^{1}$
Yin-Jie Zhang$^{2}$}

\date{}

\maketitle

\end{CJK*}

\noindent {\small 1.College of
Science,  North China University of Science and Technology, Tangshan 063210,
China}
\noindent{\small 2.College of
Physics Science and Technology,  Hebei University , Baoding 071002, China}

\baselineskip 9mm
\begin{abstract}
By globally analyzing nuclear Drell-Yan data including all incident energies, the nuclear effects of nPDFs and initial-state parton energy loss are investigated.  Based on Landau-Pomeranchuk-Migdal (LPM) regime, the calculations are carried out by means of the analytic parametrizations of quenching weights derived from the Baier-Dokshitzer-Mueller-Peign$\acute{e}$-Schiff (BDMPS) formalism and using the new EPPS16 nPDFs. It is found that the results are in good agreement with the data and the role of the energy loss effect on the suppression of Drell-Yan ratios is prominent, especially for low-mass Drell-Yan measurements. The nuclear effects of nPDFs becomes more obvious with the nuclear mass number A, the same as the energy loss effect. By global fit, the transport coefficient extracted is $\hat{q}=0.26\pm0.04$ GeV$^{2}$/fm. In addition, to avoid diminishing the QCD NLO correction on the data form of Drell-Yan ratios, the separate calculations about the Compton differential cross section ratios $R_{Fe(W)/C}(x_{F})$ at 120GeV are performed, which provides a feasible way to better distinguish the gluon energy loss in Compton scattering. It is found that the role of the initial-state gluon energy loss on the suppression of Compton scattering rations is not very important and becomes disappear with the increase of $x_{F}$.

\vskip 1.0cm



\end{abstract}

\maketitle
\newpage
\vskip 0.5cm

\section{Introduction}
The nuclear Drell-Yan production of leptons provides an ideal tool to study the parton dynamics and the nuclear effects in cold nuclear matter. The observed suppression of the nuclear Drell-Yan production ratios ($R_{A_{1}/A_{2}}$) is generally believed to be induced by the nuclear effects incorporated in nuclear parton distribution functions (nPDFs) and the initial-state parton energy loss[1-3]. The study of the Drell-Yan nuclear modification is conducive to reveal the properties of the radiative energy loss when partons propagating through the nuclear medium, and moreover, may help to better understand the energy loss effect causing the jet quenching phenomenon observed at RHIC and LHC[4,5].

In the past over three decades, NA3[6], NA10[7], E772[8], E866[9] and  E906[10] measurements have provided sufficient experiment data for studying the suppression of the Drell-Yan differential cross section ratios. Several groups have given the interpretation about the nuclear attenuation in Drell-Yan process by means of the phenomenological models based on energy loss and the nuclear effects of nPDFs [1-3, 11-14]. The Ref.[1] analyzed data from E772 and E866, and utilized a very different reference frame and prescription for calculating the shadowing . By disentangling energy loss and shadowing when analyzing experimental data, they extracted the mean quark energy loss per unit path length $dE/dz = 2.73 \pm 0.37 \pm 0.5$ GeV/fm, which is consistent with theoretical expectations including the effects of the inelastic interaction of the incident proton at the surface of the nucleus. The Ref.[12] analyzed data from E866 and NA3, and extracted the transport coefficient to be $\hat{q}=0.24\pm0.18$ GeV/fm$^{2}$ by means of EKS98 nPDFs[15] and the energy loss distribution  based on the BDMPS approach,  which corresponds to a mean energy loss per unit length $dE/dz = 0.20 \pm 0.15$ GeV/fm for $L=5$fm and $A\approx 200$.  By means of EPS08 nPDFs[16], the Ref.[14] analyzed data from E866 with the transport coefficient $\hat{q}=0.024$ GeV$^{2}$/fm (corresponds to a mean energy loss per unit length $dE/dz = 0.20$ GeV/fm for $L=5$fm ) determined from the nuclear modification of single-inclusive DIS hadron spectra as measured by the HERMES experiment[17]. Because of EKS98 nPDFs[15] and EPS08 nPDFs[16] determining nuclear shadowing of sea quarks from E866 and E772 nuclear Drell-Yan data which may be substantially contaminated by energy loss, the initial state energy loss in Drell-Yan process constrained by EKS98 or EPS08 nPDFs is underestimated due to an overestimation for nuclear shadowing correct to the sea quark distribution. So, the conclusions derived from the above two articles (Ref.[12] and Ref.[14]) are analogous.

From above comments, it can be found that the conclusions about the role of the initial-state energy loss effect on the Drell-Yan suppression is dependent on the nPDF sets used in the calculations about the Drell-Yan differential cross section ratios. Like the energy loss effect, the shadowing effect incorporated in nPDFs can also induce the suppression of Drell-Yan ratios. Since the underlying mechanisms driving the in-medium corrections of the nucleon substructure have not been completely understood, the shadowing effect has not been determined reliably. Several sets of nPDFs, such as EKS98[15], EPS08[16], EPS09[18], HKN07[19] and nDS[20], determine the distributions of the valance quark at larger momentum fraction and sea quark at smaller momentum fraction, by fitting the nuclear Drell-Yan data. This may lead to overestimate the nuclear modification in the sea quark distribution, in view of the role of energy loss on Drell-Yan suppression. Hence, by means of these sets of nPDFs, the study about the nuclear effects of nPDFs and initial-state energy loss effect in Drell-Yan process can not get a reliable conclusion. Lately, by firstly including data constraints from the new LHC experiments, neutrino DIS measurements and low-mass Drell-Yan data (NA3[6], NA10[7] and E615[21]), EPPS16[22] is given, which significantly extends the kinematic reach of the data constraints and leads to a more reliable modification about the nuclear effects of nPDFs.

The initial-state parton energy loss in nuclear Drell-Yan process is sensitive to the Landau-Pomeranchuk-Migdal (LPM) regime[23] due to the gluon formation time $t_{f}$ ($t_{f}\propto 1/q_{T}^{2}$ as expressed in Ref.[14]) being much smaller than the medium length $L_{A}$ for large values of $q_{T}$, which is different from the fully coherent energy loss (FCEL)[24]. Like initial-state (final-state) parton energy loss, FCEL can also induce a significant hadron suppression in hadron-nucleus collisions. In some hadron-nucleus collisions, these two kinds of energy loss effects both exist and it is difficult to distinguish them, such as the suppression of $J/\psi$ productions. Since FCEL is absent in nuclear Drell-Yan process, we can clearly probe the initial-state energy loss and constrain the transport coefficient in the cold nuclear medium by means of Drell-Yan measurements. This may profit to disentangle the relative contributions of the initial-state (final-state) energy loss and coherent energy loss, when they both exist in some hadron-nucleus collisions.

Up to now, the mechanism of the medium-induced parton energy loss has not been understood completely, and due to lack of the reliable determination of the nuclear PDFs or the global analysis for precision data at different incident energies, there is no consensus about the role and the transport coefficient of the initial-state parton energy loss in nuclear Drell-Yan process. In this work, by means of the new EPPS16 nPDFs[22] and the analytic parametrizations of quenching weights derived from the Baier-Dokshitzer-Mueller-Peign$\acute{e}$-Schiff (BDMPS) formalism based on LPM regime[25-27], the Drell-Yan nuclear modification due to the nuclear effects of nPDFs and initial-state quark energy loss will be investigated. To accurately extract the value of the transport coefficient, the global fit will be carried out by including all the incident energy  data from the new E906 (120GeV) to E866 (800GeV). Furthermore, at next-to-leading order (NLO), the initial-state gluon energy loss rooted in the primary NLO subprocess (Compton scattering) will be also investigated. The following is the theoretical framework expressed in Section 2, results and discussion presented in Section 3, and summary expounded in Section 4.

\section{Model for nuclear Drell-Yan suppression}
In nuclear Drell-Yan process, the incident parton undergoes multiple soft collisions accompanied by gluon emission when traveling through the nuclear medium. These radiated gluons carry away some energy $\epsilon$ of the incident parton with the probability distribution $D(\epsilon)$. In nuclear Drell-Yan hadron production, as the gluon formation time $t_{f}$ is much smaller than the medium length $L_{A}$, the parton energy loss is in the LPM regime[23] with
\begin{eqnarray}
 \langle\epsilon\rangle_{LPM}\propto\hat{q}L^{2}
 \end{eqnarray}
  where $\hat{q}$ represents the transport coefficient and $L$ is the length of traversed nuclear matter, which is different from FCEL regime[24] with
  \begin{eqnarray}
  \langle\epsilon\rangle_{FCEL}\propto\frac{\sqrt{\hat{q}L}}{M}\cdot E,
  \end{eqnarray}
  where $M$ and $E$ represent the mass and energy of the parton respectively. The formalism suitable for describing the LPM energy loss has been presented by Baier, Dokshitzer, Mueller, Peign$\acute{e}$ and Schiff (BDMPS)[25, 26].  According to the BDMPS energy loss framework, an analytic parametrization of the probability distribution $D(\epsilon)$ for LPM initial-state energy loss is derived by F. Arleo[27]:

\begin{eqnarray}
 D(\epsilon)=\frac{1}{\sqrt{2\pi}\sigma(\epsilon/\omega_{c})}\exp[-\frac{(\log(\epsilon/\omega_{c})-\mu)^{2}}{2\sigma^{2}}],
\end{eqnarray}
where $\omega_{c}=\frac{1}{2}\hat{q}L^{2}$, $\mu=-2.55$ and $\sigma=0.57$.

The energy $\epsilon$ carried away by radiated gluons results in a change about the incident parton momentum fraction prior to the hard QCD process:
\begin{eqnarray}
x_{1}\rightarrow x'_{1}=x_{1}+\epsilon/E_{beam},
\end{eqnarray}
where $x_{1}$ represents the momentum fraction of the partons in the beam hadron.
The model for LPM initial-state energy loss can be expressed as:

\begin{eqnarray}
\frac{d^{2}\sigma'_{h-A}}{dx_{F}dM^{2}}=\int_{0}^{(1-x_{1})E_{beam}}d\epsilon D(\epsilon)\frac{d^{2}\sigma_{h-A}}{dx_{F}dM^{2}}(x'_{1},x_{2},Q^{2}).
\end{eqnarray}
Here $x_{2}$ denotes the momentum fraction of the partons in the target, $x_{F}=x_{1}-x_{2}$, and the invariant mass of a lepton
pair $Q^{2}=M^{2}=sx_{1}x_{2}$ ($\sqrt{s}$ is the center of mass energy of the hadronic collision).

 The NLO Drell-Yan differential cross section consists of the partonic cross section $\frac{d^{2}\sigma^{DY}_{h-A}}{dx_{F}dM^{2}}$ from the process of the Born diagram $q\bar{q}\rightarrow \gamma^{*}$,  $\frac{d^{2}\sigma^{C}_{h-A}}{dx_{F}dM^{2}}$ from the Compton scattering $qg\rightarrow q\gamma^{*}$, $\frac{d^{2}\sigma^{Ann}_{h-A}}{dx_{F}dM^{2}}$ from the annihilation
process $q\bar{q}\rightarrow g\gamma^{*}$, and hence can be given as:
\begin{eqnarray}
\frac{d^{2}\sigma_{h-A}}{dx_{F}dM^{2}}=\frac{d^{2}\sigma^{DY}_{h-A}}{dx_{F}dM^{2}}+\frac{d^{2}\sigma^{C}_{h-A}}{dx_{F}dM^{2}}+\frac{d^{2}\sigma^{Ann}_{h-A}}{dx_{F}dM^{2}},
\end{eqnarray}
where
\begin{eqnarray}
\frac{d^{2}\sigma^{DY(C,Ann)}_{h-A}}{dx_{F}dM^{2}}=\int_{x_{1}}^{1}dt_{1}\int_{x_{2}}^{1}\frac{d^{2}\hat{\sigma}^{DY(C,Ann)}}{dx_{F}dM^{2}}\hat{Q}^{DY(C,Ann)}(t_{1},t_{2})dt_{2}.
\end{eqnarray}
Here $t_{1}$ and $t_{2}$ are the fractions of hadron momenta taken by quarks or gluons.
For the process of the Born diagram $q\bar{q}\rightarrow \gamma^{*}$:
\begin{eqnarray}
\frac{d^{2}\hat{\sigma}^{DY}}{dx_{F}dM^{2}}=\frac{4\pi\alpha^{2}}{9M^{2}s}\frac{1}{x_{1}+x_{2}}\delta(t_{1}-x_{1})\delta(t_{2}-x_{2}),
\end{eqnarray}
\begin{eqnarray}
\hat{Q}^{DY}(t_{1},t_{2})=\sum_{f}e_{f}^{2}[q_{f}^{h}(t_{1},Q^{2})\bar{q}_{f}^{A}(t_{2},Q^{2})+\bar{q}_{f}^{h}(t_{1},Q^{2})q_{f}^{A}(t_{2},Q^{2})],
\end{eqnarray}
for the Compton scattering $qg\rightarrow q\gamma^{*}$:
\begin{eqnarray}
\frac{d^{2}\hat{\sigma}^{C}}{dx_{F}dM^{2}}=\frac{3}{16}\times\frac{16\alpha^{2}\alpha_{s}(Q^{2})}{27M^{2}s}\frac{1}{x_{1}+x_{2}}C(x_{1},x_{2},t_{1},t_{2}),
\end{eqnarray}
\begin{eqnarray}
\hat{Q}^{C}(t_{1},t_{2})=\sum_{f}e_{f}^{2}\{g^{h}(t_{1},Q^{2})[q_{f}^{A}(t_{2},Q^{2})+\bar{q}_{f}^{A}(t_{2},Q^{2})]
\nonumber \\+g^{A}(t_{2},Q^{2})[q_{f}^{h}(t_{1},Q^{2})+\bar{q}_{f}^{A}(t_{1},Q^{2})]\},
\end{eqnarray}
and for the annihilation process $q\bar{q}\rightarrow g\gamma^{*}$:
\begin{eqnarray}
\frac{d^{2}\hat{\sigma}^{Ann}}{dx_{F}dM^{2}}=\frac{1}{2}\times\frac{16\alpha^{2}\alpha_{s}(Q^{2})}{27M^{2}s}\frac{1}{x_{1}+x_{2}}Ann(x_{1},x_{2},t_{1},t_{2}),
\end{eqnarray}
\begin{eqnarray}
\hat{Q}^{Ann}(t_{1},t_{2})=\sum_{f}e_{f}^{2}[q_{f}^{h}(t_{1},Q^{2})\bar{q}_{f}^{A}(t_{2},Q^{2})+\bar{q}_{f}^{h}(t_{1},Q^{2})q_{f}^{A}(t_{2},Q^{2})].
\end{eqnarray}
In the above formulas, $q^{h(A)}_{f}(x_{1(2)},Q^{2})$ refers to the parton distribution function with flavor $f$ in the hadron (nucleus A), $e_{f}$ denotes the charge
of the quark with flavor $f$, $\alpha$ represents the fine structure constant, $\alpha_{s}(Q^{2})$ is the specific expression of the function, the complex expressions of functions $C(x_{1},x_{2},t_{1},t_{2})$ and $Ann(x_{1},x_{2},t_{1},t_{2})$ are showed in the Ref.[28]. The partonic densities of the nucleus A is different from that of a free proton due to the complex nuclear environment, such as  EMC suppression, shadowing and anti-shadowing. In this paper, the Drell-Yan nuclear modification due to the nuclear effects of nPDFs is computed with the latest EPPS16 set[22].

In the above energy loss correction model for interpreting the nuclear Drell-Yan suppression,  the transport coefficient $\hat{q}$  is the only parameter, which  measures the properties of the initial-state energy loss effect in cold medium and can be constrained from $\chi^{2}$ analysis about the fit of the experimental data by calculating the Drell-Yan differential cross section ratio:

\begin{eqnarray}
R_{A_{1}/A_{2}}(x_{F})=\frac{A_{2}}{A_{1}}(\frac{d^{2}\sigma'_{h-A1}}{dx_{F}dM}/\frac{d^{2}\sigma'_{h-A2}}{dx_{F}dM})
\end{eqnarray}

\section{Results and discussion}
Firstly, we investigate the nuclear effects of nPDFs on the nuclear Drell-Yan ratio using the EPPS16 nPDFs[22] together with nCTEQ15 parton density of the proton[29] or the parton distributions of the negative pion[30]. It should be noted that in our calculation the corrections from isospin effects are neglected due to their small influence on Drell-Yan ration[2]. The solid lines in Fig.1, 2, 3, 4, 5 show the next-to-leading order Drell-Yan ratios $R_{A_{1}/A_{2}}$ and the dashed lines are corresponding to the leading order calculations ($q\bar{q}\rightarrow \gamma^{*}$), and they are both modified only by the nPDFs corrections. It is found that the theoretical results at next-to-leading order and leading order are almost identical for E906, NA3 and NA10 experiment, and there is a small difference between the two results for E866 experiments. From the above expression of $\frac{d^{2}\sigma^{DY}_{h-A}}{dx_{F}dM^{2}}$ (see Eq.9),  $\frac{d^{2}\sigma^{C}_{h-A}}{dx_{F}dM^{2}}$ (see Eq.11) and $\frac{d^{2}\sigma^{Ann}_{h-A}}{dx_{F}dM^{2}}$ (see Eq.13) in Section 2, we can calculate and derive that:
\begin{eqnarray}
\frac{d^{2}\sigma^{DY}_{h-A_{1}}}{dx_{F}dM^{2}}/\frac{d^{2}\sigma^{DY}_{h-A_{2}}}{dx_{F}dM^{2}}\approx\frac{d^{2}\sigma^{Ann}_{h-A_{1}}}{dx_{F}dM^{2}}/\frac{d^{2}\sigma^{Ann}_{h-A_{2}}}{dx_{F}dM^{2}}
\approx\frac{d^{2}\sigma^{C}_{h-A_{1}}}{dx_{F}dM^{2}}/\frac{d^{2}\sigma^{C}_{h-A_{2}}}{dx_{F}dM^{2}},
\end{eqnarray}
in the momentum fraction range that the nuclear effects of gluon distributions are not gigantic. Therefore, the form of the differential cross section ratio given by the nuclear Drell-Yan data actually diminishes the QCD next-to-leading order correction. From Fig.1, 2, 3, 4, 5, we can also see that there is an obvious deviation between the calculations obtained by only including the EPPS16 nPDFs corrections and the measurements of E906 (120GeV), NA3 (150GeV) as well as NA10 (140GeV) experiments at the lower incident energies, However, a good fit can be seen between the results and the E866 (800GeV) as well as NA10 (286GeV) measurements with the higher beam energies. Further, in Table 1, we compute the $\chi^{2}/N$ (N is the number of data points) at leading order calculation. The exhibitions of Table 1 also show that the nuclear effects of nPDFs play more important role on the Drell-Yan nuclear modification with the increase of the beam energy.

\begin{figure}
\centering
\includegraphics*[width=18cm, height=12cm]{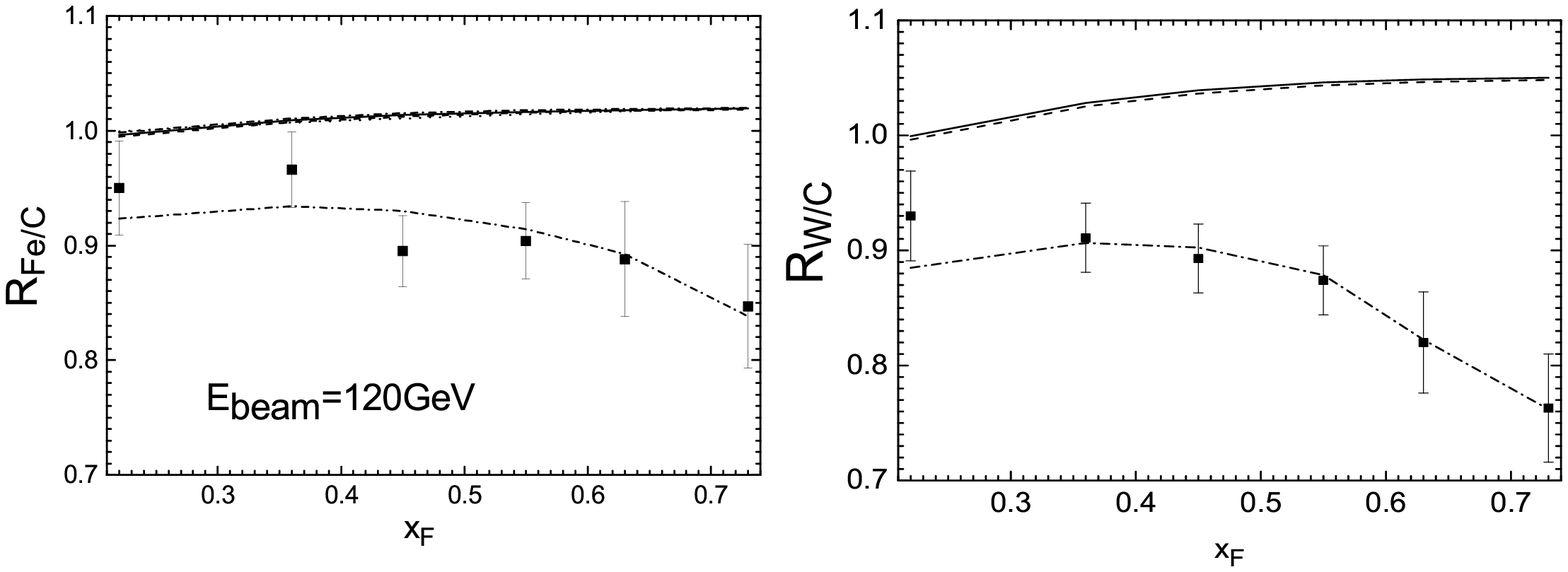}
\vspace{-6.5cm} \caption{E906 nuclear Drell-Yan ratios $R_{Fe/C}(x_{F})$ (left) and $R_{W/C}(x_{F})$ (right) compared to the theoretical results with  EPPS16 nPDFs at next-to-leading order (solid lines), leading order calculation (dashed lines), and considering initial-state quark energy loss from the process of the Born diagram $q\bar{q}\rightarrow \gamma^{*}$ (dashed-dotted lines).  }
\vspace{-0.1cm}
\end{figure}

\setlength{\abovecaptionskip}{-0.2cm}
\begin{figure}
\centering
\includegraphics*[width=18cm, height=12cm]{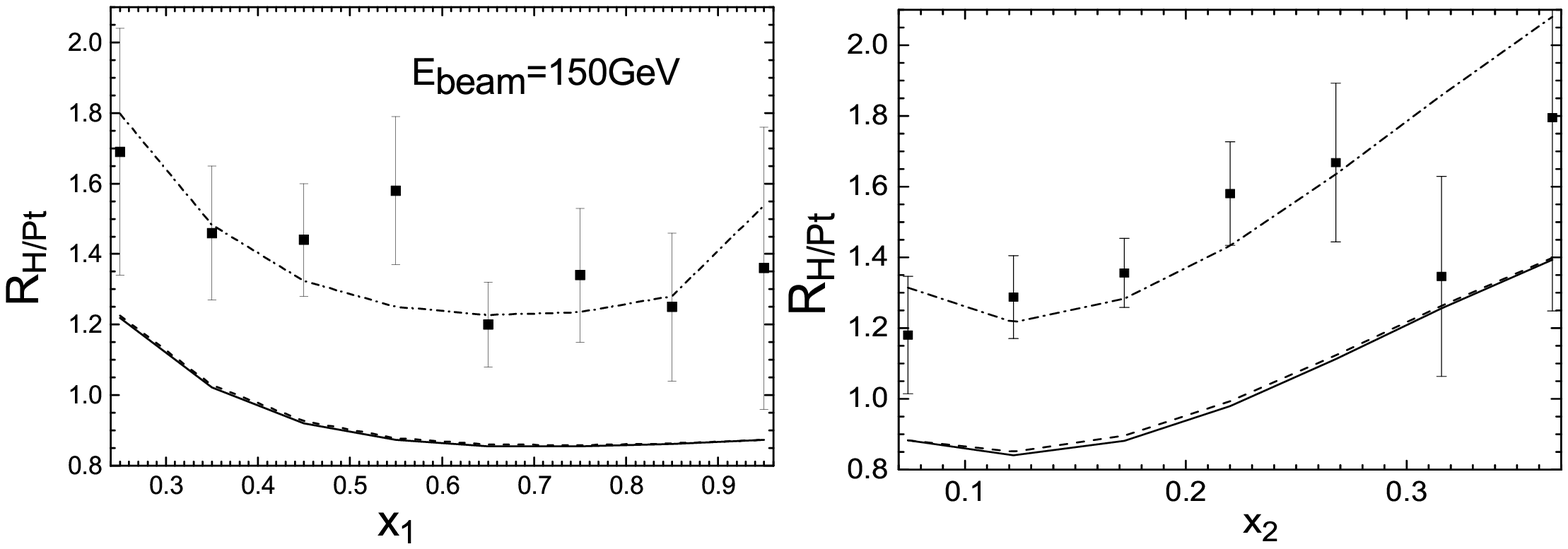}
\vspace{-6.5cm} \caption{NA3 nuclear Drell-Yan ratios $R_{H/Pt}(x_{1})$ (left) and $R_{H/Pt}(x_{2})$ (right) compared to the theoretical results with  EPPS16 nPDFs at next-to-leading order (solid lines), leading order calculation (dashed lines), and considering initial-state quark energy loss from the process of the Born diagram $q\bar{q}\rightarrow \gamma^{*}$ (dashed-dotted lines). }
\vspace{-0.8cm}
\end{figure}

\setlength{\abovecaptionskip}{-0.2cm}
\begin{figure}
\centering
\includegraphics*[width=26cm, height=18cm]{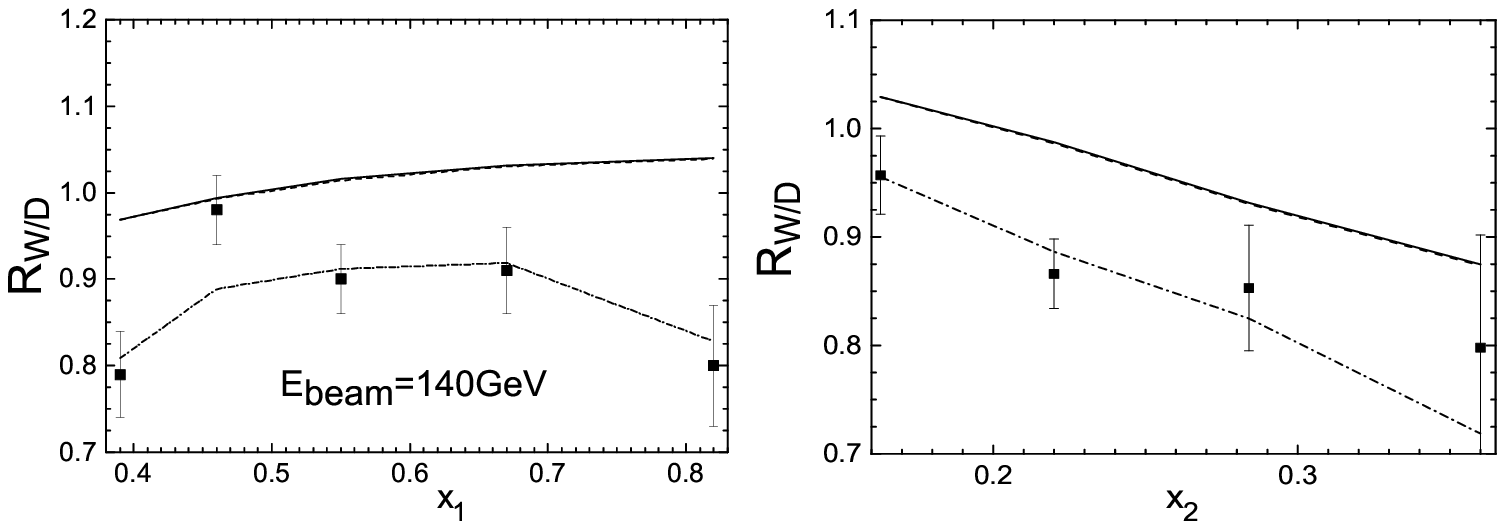}
\vspace{-12.5cm} \caption{NA10(140GeV) nuclear Drell-Yan ratios $R_{W/D}(x_{1})$ (left) and $R_{W/D}(x_{2})$ (right) compared to the theoretical results with  EPPS16 nPDFs at next-to-leading order (solid lines), leading order calculation (dashed lines), and considering initial-state quark energy loss from the process of the Born diagram $q\bar{q}\rightarrow \gamma^{*}$ (dashed-dotted lines). }
\vspace{-0.8cm}
\end{figure}

\setlength{\abovecaptionskip}{-0.2cm}
\begin{figure}
\centering
\includegraphics*[width=26cm, height=20cm]{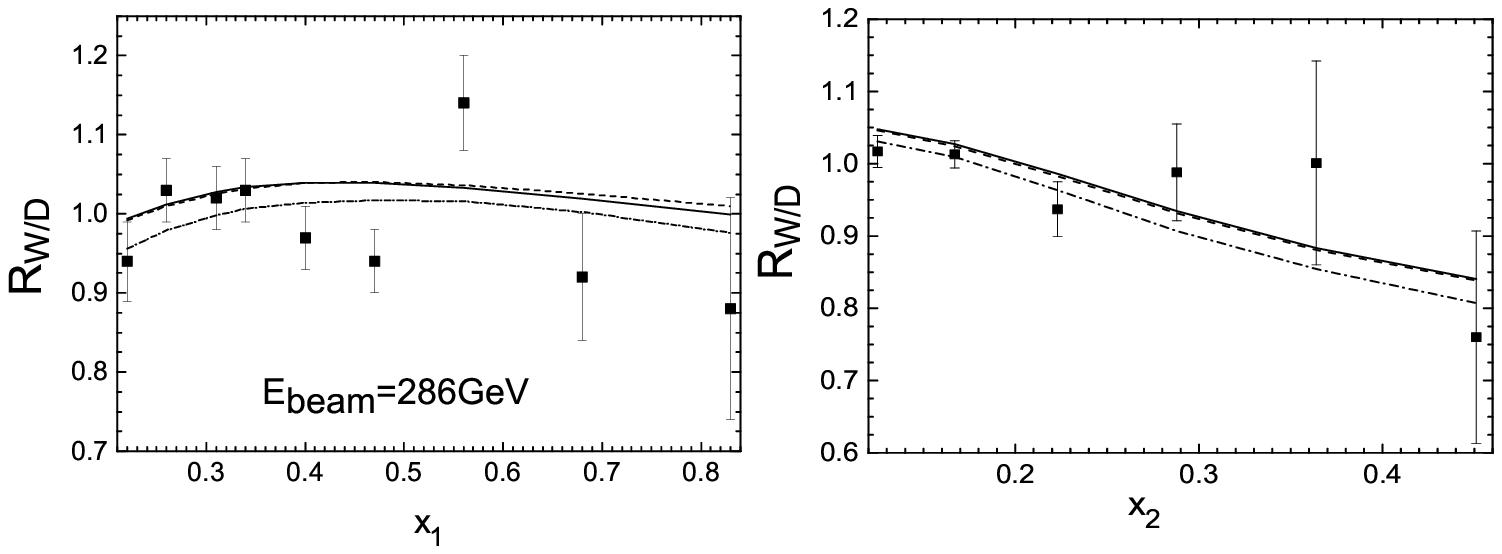}
\vspace{-13cm} \caption{NA10(286GeV) nuclear Drell-Yan ratios $R_{W/D}(x_{1})$ (left) and $R_{W/D}(x_{2})$ (right) compared to the theoretical results with  EPPS16 nPDFs at next-to-leading order (solid lines), leading order calculation (dashed lines), and considering initial-state quark energy loss from the process of the Born diagram $q\bar{q}\rightarrow \gamma^{*}$ (dashed-dotted lines). }
\vspace{-0.8cm}
\end{figure}

\setlength{\abovecaptionskip}{-0.2cm}
\begin{figure}
\centering
\includegraphics*[width=18cm, height=12cm]{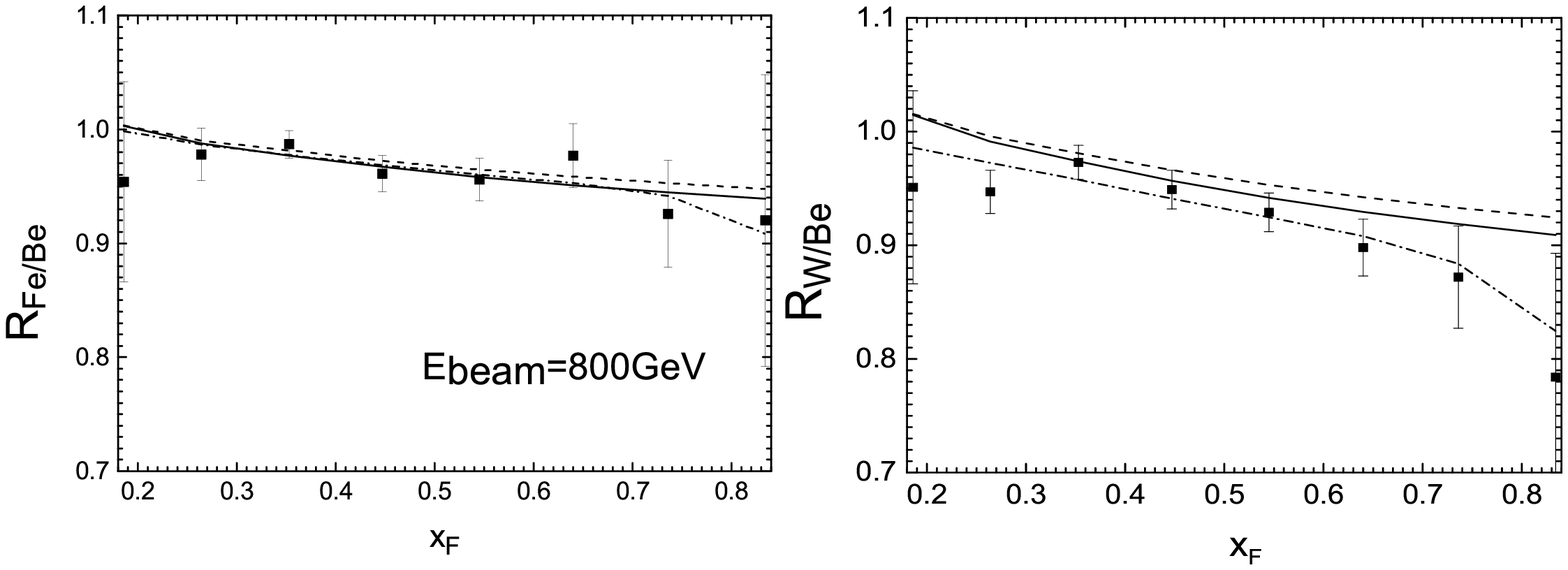}
\vspace{-6cm} \caption{E866 nuclear Drell-Yan ratios $R_{Fe/Be}(x_{F})$ (left) and $R_{W/Be}(x_{F})$ (right) compared to the theoretical results with  EPPS16 nPDFs at next-to-leading order (solid lines), leading order calculation (dashed lines), and considering initial-state quark energy loss from the process of the Born diagram $q\bar{q}\rightarrow \gamma^{*}$ (dashed-dotted lines). }
\vspace{-0.8cm}
\end{figure}

\begin{table}
\caption{The $\chi^{2}/N$ values obtained only with EPPS16 nPDFs[22].}
\begin{center}
\begin{tabular}{p{5cm}p{3cm}p{3cm}p{3cm}c}\hline
Exp.data   &Data points & Beam &$\chi^{2}/ndf$  \\
\hline
E906($x_{F}$)& 12 &120GeV   & 15.07   \\
NA3($x_{1(2)}$)  & 15 &150GeV   & 7.31   \\
NA10($x_{1(2)}$)& 9 &140Gev   & 6.55  \\
NA10($x_{1(2)}$)& 15 &286GeV   & 1.43  \\
 E866($x_{F}$)   & 16 &800GeV    & 1.22   \\
\hline
\end{tabular}
\end{center}
\end{table}


\begin{table}
\caption{The $\hat{q}$ and $\chi^{2}/ndf$ values obtained with EPPS16 nPDFs[22] and initial-state quark energy loss.}
\begin{center}
\begin{tabular}{p{5cm}p{1.6cm}p{4cm}p{2.4cm}p{1.4cm}c}\hline
Exp.data   &Data points & Momentum fraction  &$\hat{q}$(GeV$^{2}$/fm)&$\chi^{2}/ndf$  \\
\hline
E906($R_{Fe/C}(x_{F})$)& 6 & $0.22<x_{F}<0.73$  & $0.45\pm0.06$ &0.46 \\
E906($R_{W/C}(x_{F})$)& 6 & $0.22<x_{F}<0.73$  & $0.25\pm0.02$ &0.25 \\
Glob fit E906-120GeV & 12 &   & $0.25\pm0.02$ &0.91 \\
\hline
NA3($R_{H/Pt}(x_{1})$)  & 8 &$0.25<x_{1}<0.95$ & $0.24\pm0.11$ &0.46  \\
NA3($R_{H/Pt}(x_{2})$)  & 7 &$0.074<x_{2}<0.366$   & $0.24\pm0.11$ &0.88  \\
Glob fit NA3-150GeV  & 15 &  & $0.24\pm0.10$ &0.65  \\
\hline
NA10-140GeV($R_{W/D}(x_{1})$)& 5 & $0.39<x_{1}<0.82$  & $0.35\pm0.02$ &1.13  \\
NA10-140GeV($R_{W/D}(x_{2})$)& 4 & $0.163<x_{2}<0.360$     & $0.27\pm0.04$ &0.31  \\
Glob fit NA10-140GeV& 9 &  & $0.30\pm0.05$  &0.99\\
\hline
NA10-286GeV($R_{W/D}(x_{1})$)& 9 & $0.22<x_{1}<0.83$  & $0.19\pm0.11$ &1.45  \\
NA10-286GeV($R_{W/D}(x_{2})$)& 6 & $0.125<x_{2}<0.451$     & $0.10\pm0.07$ &0.59  \\
Glob fit NA10-286GeV& 15 &  & $0.14\pm0.05$  &1.14\\
\hline
 E866($R_{Fe/Be}(x_{F})$)   & 8 & $0.186<x_{F}<0.834$     & $0.17\pm0.17$ &0.27   \\
 E866($R_{W/Be}(x_{F})$)   & 8 & $0.186<x_{F}<0.834$     & $0.39\pm0.10$ &0.46   \\
  Glob fit E866-800GeV   & 16 &    & $0.36\pm0.10$ &0.41   \\
 \hline
Global fit  &67& & $0.26\pm0.04$&0.82\\

\hline
\end{tabular}
\end{center}
\end{table}

Secondly, we appreciate the initial-state quark energy loss in nuclear Drell-Yan production. In view of the correction model for initial-state energy loss and the above discussion, the energy loss effect in the Compton scattering and annihilation processes also can be diminished due to the form of the differential cross section ratio given by the nuclear Drell-Yan data. Therefore, in a leading order calculation, only the quark energy loss from the process of the Born diagram $q\bar{q}\rightarrow \gamma^{*}$ is considered. We calculate Drell-Yan ratios $R_{A_{1}/A_{2}}$ using the EPPS16 nPDFs[22] together with the analytic parametrization of quenching weights based on BDMPS formalism[27](Eq.3). The values of transport coefficient $\hat{q}$ and $\chi^{2}/ndf$ extracted from the corresponding  experimental data are showed in Table 2. The comparison between Table 2 and Table 1 displays that, when considering the initial-state quark energy loss effect, the fitting degree of the calculation results with the experimental data is greatly improved, especially for low incident energy data (E906-120GeV, NA3-150GeV and NA10-140GeV).

The plot on $\chi^2$ as a function of $\hat{q}$ for the global fit of all data is given in Fig.6. From Fig.6, we can easily and clearly see that the global fit shows the best value is $\hat{q}=0.26\pm0.04$ GeV$^{2}$/fm ($\chi^{2}/ndf=0.82$), which is a little smaller than the result $\hat{q}=0.32\pm0.04$ GeV$^{2}$/fm obtained by HKM nPDFs[31] in our previous work[3]. Since HKM nPDFs[31] is obtained only using the experimental data about nuclear structure functions, the shadowing effect of nPDF on the suppression of Drell-Yan ratios has been reduced in the $0.01 < x < 0.3 $ region.  In addition, as discussed in Ref.[12] the mean BDMPS energy loss $\langle\epsilon\rangle$ experienced by the fast parton in the medium is given by:
\begin{eqnarray}
\langle\epsilon\rangle\equiv \int D(\epsilon)d\epsilon=\frac{1}{2}\alpha C_{R}\omega_{c}
\end{eqnarray}
($\alpha=\frac{1}{2}, C_{R}=\frac{4}{3} $) and the mean quark energy loss per unit path length $dE/dz$ is:
\begin{eqnarray}
\frac{dE}{dZ}\equiv\frac{\langle\epsilon\rangle}{L}=\delta\times(\frac{L}{10fm})
\end{eqnarray}
($\delta$ is the parameter simply related to the transport coefficient $\hat{q}$ and means the mean BDMPS energy loss per unit path length with $L=10$fm), it can be derived that the mean BDMPS energy loss per unit path length $dE/dz=\frac{1}{6}\hat{q}L$.
Here the transport coefficient $\hat{q}=0.26\pm0.04$ GeV$^{2}$/fm corresponds to $\frac{dE}{dZ}\approx1.10\pm0.17$ GeV/fm for $L=5$fm, which is much bigger
than the result $\frac{dE}{dZ}=0.20\pm0.15$ GeV/fm obtained by using EKS98 nPDFs in Ref.[12], the result $\frac{dE}{dZ}=0.20$ GeV/fm obtained by EPS08 nPDFs in Ref.[14], or the result $\frac{dE}{dZ}=0.23\pm0.09$ GeV/fm obtained by EPS09 nPDFs in Ref.[13] . For $L=10$fm, the transport coefficient $\hat{q}=0.26\pm0.04$ GeV$^{2}$/fm corresponds to $\frac{dE}{dZ}\approx2.20\pm0.34$ GeV/fm, which is approximately identical with the result $\frac{dE}{dZ}=2.73\pm0.37\pm0.5$ GeV/fm obtained by unambiguously separating shadowing and energy loss in Ref.[1]. This indicates that with EPPS16 nPDFs the value of quark energy loss actually can be better constrained from the nuclear Drell-Yan data by avoiding overestimating the shadowing correction. The reason is that although the fit of EPPS16 nPDFs includes the nuclear Drell-Yan data, the new data (NA3[6], NA10[7] and E615[21]) with respect to the EPS09 increase the variety of the momentum fraction of the target parton from 0.074 to 0.451 and efficiently avoid overestimating the nuclear modification about the sea quark distribution. Furthermore, the fit of EPPS16 nPDFs firstly includes the data from higher energy of LHC proton-lead collisions, which can completely disregard energy loss and provide better constraints for the A dependence of the parton nuclear modifications. For large beam energy such as E866, E772 nuclear Drell-Yan experiments, the large portion of data comes with $x_{2} < 0.05$, which falls in the region of significant nuclear shadowing. For low beam energy such as  NA3, NA10 and E906 measurements,  the large portion of data comes with  $0.1 <x_{2} < 0.45$, which falls in the region of only tiny (anti-)shadowing.  The nuclear modification of Drell-Yan process is sensitive to nPDFs mainly due to some kinds of nPDFs (such as EPS09, EPS08 and EKS98) determining nuclear shadowing of sea quarks from E866 and E772 nuclear Drell-Yan data which may be substantially contaminated by energy loss. To minimize the dependence on nPDFs, the nuclear Drell-Yan measurements at lower beam energy should provide better constraints for the initial-state parton energy loss.

\begin{figure}
\centering
\includegraphics*[width=10cm, height=8cm]{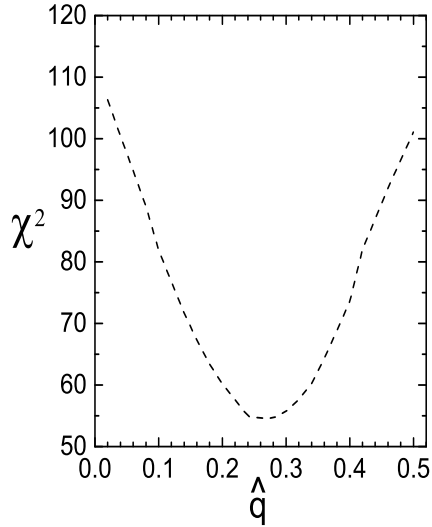}
\vspace{-0cm} \caption{The plot on $\chi^2$ as a function of $\hat{q}$ for the global fit of all data. }
\end{figure}

With the value of the transport coefficient $\hat{q}$ displayed in Table 2, the results considering the energy loss effect from the process of the Born diagram $q\bar{q}\rightarrow \gamma^{*}$ are showed as the dashed-dotted lines in Fig.1, 2, 3, 4, 5. It is found that the calculations by EPPS16 nPDFs
together with the initial-state quark energy loss effect are in good agreement with the Drell-Yan data, especially for low-mass Drell-Yan measurements (E906-120GeV, NA3-150GeV and  NA10-140GeV), and the fitting degree is better than the results acquired with using EPPS16 nPDF and the fully coherent regime $\hat{q}=0.07-0.09$ GeV$^{2}$/fm extracted from $J/\psi$ measurements[2]. It also displays that the role of the incident quark energy loss effect on the suppression of Drell-Yan ratios reduces with the increase of the beam energy, and becomes more important with the increase of the nuclear mass number A.

\begin{figure}[ht]
\centering
\includegraphics*[width=18cm, height=12cm]{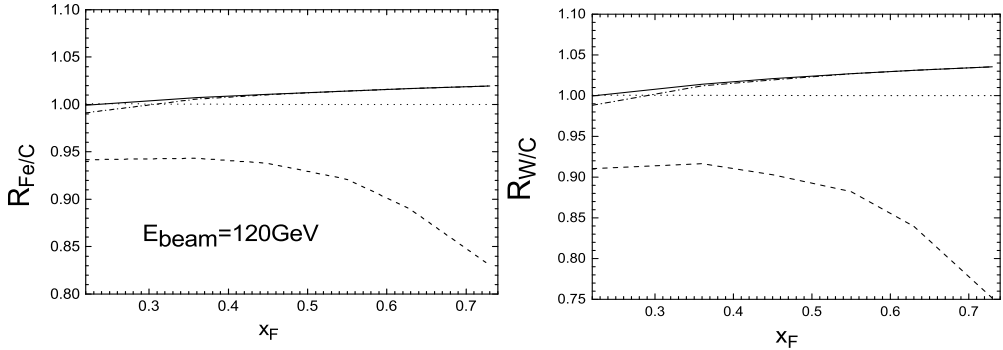}
\vspace{-6cm} \caption{The differential cross section ratios $R_{Fe/C}(x_{F})$ (left) and $R_{W/C}(x_{F})$ (right) from the Compton scattering $qg\rightarrow q\gamma^{*}$.
The dotted (solid) lines represent the results without (with) the nuclear modification of the gluon distribution, and the dashed-dotted (dashed) lines correspond to the calculations with adding the initial-state energy loss of gluon (quark and gluon).  }
\end{figure}

Thirdly, we appraise the role of the initial-state gluon energy loss by means of the primary NLO subprocess (Compton scattering) in Drell-Yan production. At next-to-leading order, the differential cross section of Compton scattering $qg\rightarrow q\gamma^{*}$ involves the gluon distributions of the incident hadron and the target nucleus, as can be seen in Eq.11. This provides an opportunity to explore the initial-state gluon energy loss effect.  In order to avoid diminishing the QCD NLO correction on the data form of Drell-Yan ratios and better investigate the gluon energy loss in Compton scattering $qg\rightarrow q\gamma^{*}$, we separately calculate the Compton differential cross section ratios $R_{Fe(W)/C}(x_{F})$ at 120GeV. The obtained results can be seen in Fig.7. The dotted lines represent the results without the nuclear modification about the gluon distribution of the target nucleus, the solid lines include the gluon nuclear effects, the dashed-dotted lines correspond to the calculations with adding the initial-state energy loss of gluon, and the dashed lines include both quark and gluon energy loss. From Fig.7, it is can be seen that the deviation between the dotted lines and the solid lines increases to approximately $2\%$ for $R_{Fe/C}(x_{F})$ and $3.5\%$ for $R_{W/C}(x_{F})$ at $x_{F}\approx0.73$, which indicates that the role of the nuclear effects of gluon distributions in Compton scattering is apparent and becomes more important with the increase of $x_{F}$ as well as the nuclear mass number A. The deviation between the solid and the dashed-dotted lines is approximately from $1\%$ to 0 with $x_{F}$ from 0.22 to 0.73, which illustrates that the role of the initial-state gluon energy loss effect on the suppression of Compton scattering rations is not very important, and reduces with the increase of $x_{F}$. The deviation between the dashed-dotted lines and the dashed lines is approximately from $5\%$ to $19\%$ for $R_{Fe/C}(x_{F})$ and from $8\%$ to $29\%$ for $R_{W/C}(x_{F})$ with $x_{F}$ from 0.22 to 0.73, which illustrates that the initial-state quark energy loss effect in Compton scattering becomes more significant with the increase of $x_{F}$ and the nuclear mass number A. It is clear, in the range $0.22<x_{F}<0.73$ at $E_{beam}=120$GeV, the initial-state quark energy loss is the dominant effect which induces the suppression of Compton scattering rations, the nuclear effects of gluon leading to the rise of Compton scattering rations are obvious, and the initial-state gluon energy loss has an influence on the suppression in small $x_{F}$. This means that it may be feasible for investigating the initial-state gluon energy loss from the separate calculation about the primary NLO subprocess (Compton scattering) in Drell-Yan production at small $x_{F}$ range and lower incident energy.

\section{ Summary }
By means of the new EPPS16 nPDFs[22] and the analytic parametrizations of quenching weights derived from the BDMPS formalism based on LPM regime[25-27], the Drell-Yan nuclear modification due to the nuclear effects of nPDFs and initial-state parton energy loss is investigated, by globally analyzing the all incident energy experimental data (67 points) including E906-120GeV[10], NA3-150GeV[6], NA10-140GeV[7], NA10-286GeV[7], and E866-800GeV[9]. It is found that the calculations by EPPS16 nPDF together with the initial-state energy loss effect are in good agreement with the Drell-Yan data and the role of the energy loss effect on the suppression of Drell-Yan ratios is prominent, especially for low-mass Drell-Yan measurements. The nuclear effects of nPDFs become more obvious with the nuclear mass number A, the same as the energy loss effect. The values of transport coefficient extracted by global fit is $\hat{q}=0.26\pm0.04$ GeV$^{2}$/fm ($\chi^{2}/ndf=0.82$).

In addition, to avoid diminishing the QCD NLO correction on the data form of Drell-Yan ratios and better investigate the gluon energy loss in Compton scattering $qg\rightarrow q\gamma^{*}$, we separately calculate the Compton differential cross section ratios $R_{Fe(W)/C}(x_{F})$ at 120GeV. The calculations indicate that the nuclear effects of gluon distributions leading to the rise of Compton scattering rations are obvious and become more important with the increase of $x_{F}$ and the nuclear mass number A, the role of the initial-state gluon energy loss on the suppression of Compton scattering rations is not very important and becomes disappear with the increase of $x_{F}$ and the nuclear mass number A, and the initial-state quark energy loss effect is the dominant effect which induces the suppression of Compton scattering rations and becomes more significant with the increase of $x_{F}$. This mean that it may be feasible for investigating the energy loss of gluon from the separate calculation about the primary NLO subprocess (Compton scattering) in Drell-Yan production at small $x_{F}$ range and lower incident energy.



\begin{thebibliography}{00}
\bibitem{s1} M.B.Johnson etal., Phys. Rev. C65, 025203 (2002).
\bibitem{s2} F. Arleo, C.-J. Na$\ddot{i}$m and S. Platchkov, JHEP1901, 129 (2019).
\bibitem{s3} L.-H. Song and L.-W. Yan, Phys. Rev. C 96, 045203 (2017).
\bibitem{s4} N. Armesto and E. Scomparin, Eur. Phys. J. Plus131, 52 (2016).
\bibitem{s5} G.-Y. Qin and X.-N. Wang, Int. J. Mod. Phys. E24, 1530014 (2015).
\bibitem{s6} J. Badier et al., Phys. Lett. B104, 335 (1981).
\bibitem{s7} P. Bordal et al., Phys. Lett. B193, 368 (1987).
\bibitem{s8} D. M. Alde et al., Phys.Rev.Lett. 64, 2479 (1990).
\bibitem{s9} M. A. Vasiliev et al., Phys.Rev.Lett. 83, 2304 (1999).
\bibitem{s10} P.-J. Lin, http://lss.fnal.gov/archive/thesis/2000/fermilab-thesis-2017-18.pdf,
Ph.D. thesis, Colorado U., 2017. 10.2172/1398791.
\bibitem{s11} G. T. Garvey and J. C. Peng, Phys. Rev. Lett. 90, 092302 (2003).
\bibitem{s12} F. Arleo, Phys.Lett.B 532, 231(2002).
\bibitem{s13} L.-H. Song and C.-G. Duan, Phys. Lett. B708, 68 (2012).
\bibitem{s14} Hongxi Xing et al., Nucl.Phys.A 879, 77 (2012).
\bibitem{s15} K. J. Eskola, V. J. Kolhinen, and P. V. Ruuskanen, Nucl. Phys.
B 535, 351 (1998).
\bibitem{s16} K. J. Eskola, H. Paukkunen, C. A. Salgado. JHEP 0807, 102 (2008).
\bibitem{s17} W. T. Deng, X. N. Wang, Phys. Rev. C81, 024902 (2010).
\bibitem{s18} K. J. Eskola, H. Paukkunen, and C. A. Salgado, J. High Energy
Phys. 04 (2009) 065.
\bibitem{s19} M. Hirai, S. Kumano, T.H. Nagai, Phys. Rev. C 76, 065207(2007).
\bibitem{s20} D. de Florian, R. Sassot, Phys. Rev. D 69, 074028(2004).
\bibitem{s21} J. G. Heinrich et al., Phys. Rev. Lett. 63, 356(1989).
\bibitem{s22} K. J. Eskola, P. Paakkinen, H. Paukkunen and C. A. Salgado, Eur. Phys. J. C77, 163 (2017).
\bibitem{s23} S. Peign$\acute{e}$ and A. Smilga, Phys. Usp.52, 659 (2009).
\bibitem{s24} F.Arleo, R.Kolevatov and S.Peign$\acute{e}$, Phys. Rev.D93,014006 (2016).
\bibitem{s25} R. Baier etal., Nucl.Phys. B484, 265 (1997).
\bibitem{s26} R. Baier et al., High Energy Phys. 09, 033 (2001).
\bibitem{s27} F. Arleo, J. High Energy Phys. 11, 044 (2002).
\bibitem{s28} J.Kubar et al., Nucl.Phys. B175, 251 (1980).
\bibitem{s29} K. Kovarik et al., Phys. Rev. D93, 085037 (2016).
\bibitem{s30} M. Gl$\ddot{u}$ck, E. Reya and I. Schienbein, Eur. Phys. J. C10, 313 (1999).
\bibitem{s31} M. Hirai, S. Kumano, M. Miyama, Phys. Rev. D 64, 034003(2001).

\end{thebibliography}
\end{document}